\documentclass[twocolumn,prl,superscriptaddress]{revtex4}

\usepackage{dcolumn}
\usepackage{amsmath}
\usepackage{calrsfs}

\usepackage{graphicx}
\usepackage{subfigure}

\newcommand{\set}[1]{\lbrace#1\rbrace}
\newcommand{\av}[1]{\left\langle#1\right\rangle}
\newcommand{\etal}{{\it{}et~al.}}
\newcommand{\defn}{\textit}

\newcommand{\mat}{\mathbf}
\renewcommand{\vec}{\mathbf}
\newcommand{\from}{\leftarrow}

\begin{document}

\title{Percolation on sparse networks}
\author{Brian Karrer}
\affiliation{Department of Physics, University of Michigan, Ann Arbor,
  MI 48109, USA}
\author{M. E. J. Newman}
\affiliation{Department of Physics, University of Michigan, Ann Arbor,
  MI 48109, USA}
\author{Lenka Zdeborov\'a}
\affiliation{Institut de Physique Th\'eorique, CEA Saclay and URA 2306,
  CNRS, Gif-sur-Yvette, France}

\begin{abstract}
  We study percolation on networks, which is used as a model of the
  resilience of networked systems such as the Internet to attack or failure
  and as a simple model of the spread of disease over human contact
  networks.  We reformulate percolation as a message passing process and
  demonstrate how the resulting equations can be used to calculate, among
  other things, the size of the percolating cluster and the average cluster
  size.  The calculations are exact for sparse networks when the number of
  short loops in the network is small, but even on networks with many short
  loops we find them to be highly accurate when compared with direct
  numerical simulations.  By considering the fixed points of the message
  passing process, we also show that the percolation threshold on a network
  with few loops is given by the inverse of the leading eigenvalue of the
  so-called non-backtracking matrix.
\end{abstract}

\maketitle

Percolation, the random occupation of sites or bonds on a lattice or
network with independent probability~$p$, is one of the best-studied
processes in statistical physics.  It is used as a model of porous
media~\cite{Machta91,MG95}, granular and composite
materials~\cite{OT98,Tobochnik99,BFD92,BG92}, resistor
networks~\cite{ARC85}, forest fires~\cite{Henley93}, and many other systems
of scientific interest.  In this paper we study the bond (or edge)
percolation process on general networks or graphs, which is used to model
the spread of disease~\cite{Grassberger83,Newman02c} and network
robustness~\cite{CEBH00,CNSW00,Holme02a} in social and technological
networks, among other things.  Although percolation has been studied
extensively on simple model networks such as random
graphs~\cite{CEBH00,CNSW00,GDM08,Janson09}, there are few analytic results
for real-world networks, whose structure is typically more complicated.  We
show that percolation properties of networks can be calculated using a
message passing technique, leading to a range of new results.  In
particular, we derive equations for the size of the percolating cluster and
the average size of non-percolating clusters, which can be solved rapidly
by numerical iteration given the structure of a network and the value
of~$p$.  By expanding the message passing equations about the critical
point we also derive an expression for the position of the percolation
threshold, showing that the critical value of~$p$ is given by the inverse
of the leading eigenvalue of the so-called non-backtracking
matrix~\cite{Hashimoto89,Krzakala13}, an edge-based matrix representation
of network structure that has found recent use in studies of community
detection and centrality in networks~\cite{Krzakala13,MZN14}.  The
quantities we calculate are averages over all possible realizations of the
randomness inherent in the percolation process, rather than over a single
realization, obviating the need for a separate average over realizations as
is typically required in direct numerical simulations.

We focus in particular on sparse networks, those for which only a small
fraction of possible edges are present, which includes most real-world
networks.  Our results are exact for large, sparse networks that contain a
vanishing density of short loops, but even for networks that do contain
loops, as most real-world networks do, we find the cluster size
calculations to be highly accurate and the threshold calculations can be
shown to give a lower bound on the true percolation threshold.

Consider, then, a bond percolation process on an arbitrary undirected
network of $n$ nodes and $m$ edges.  Edges are occupied uniformly at random
with probability~$p$ or unoccupied with probability~$1-p$.  The primary
entities of interest are the percolation clusters, sets of nodes connected
by occupied edges.  Since percolation is a random process, one cannot know
with certainty the identity of the clusters ahead of time, but some things
are known.  In general there will (with high probability) be at most one
percolating cluster, a cluster that fills a non-vanishing fraction of the
network in the limit of large~$n$, plus an extensive number of small
clusters of finite average size.  The percolating cluster appears only for
sufficiently large values of~$p$ and the percolation threshold~$p_c$ is the
value above which it appears; below $p_c$ there are only small clusters.

Define~$\pi_i(s)$ to be the probability that node~$i$ belongs to a small
cluster of exactly~$s$ nodes, averaged over many realizations of the random
percolation process.  If the network is a perfect tree---if it contains no
loops---then the size~$s$ of the cluster is equal to one (for node~$i$
itself) plus the sum of the numbers of nodes reachable along each edge
attached to~$i$, which is zero if the edge is unoccupied or nonzero
otherwise.  If, on the other hand, there are loops in the network then this
calculation will not, in general, give the exact value of~$s$, since it may
be possible to reach the same node along two different occupied edges,
which leads to overcounting.  If the network is sparse, however, and
\textit{locally tree-like}, meaning that in the limit of large network size
an arbitrarily large neighborhood around any node takes the form of a tree
(and hence contains no loops), then our calculation gives a good
approximation, which becomes exact in the $n\to\infty$ limit.

Working in the large~$n$ limit then and assuming the network to be locally
tree-like, the probability~$\pi_i(s)$ is equal to the probability that the
numbers of nodes reachable along each edge from~$i$ add up to~$s-1$, which
we can write as
\begin{equation}
\pi_i(s) = \sum_{\set{s_j:j\in\mathcal{N}_i}}
           \Biggl[ \prod_{j\in\mathcal{N}_i} \pi_{i\from j}(s_j) \Biggr]
  \,\delta\biggl( s-1, \sum_{j\in\mathcal{N}_i} s_j \biggr),
\end{equation}
where $\pi_{i\from j}(s)$ is the probability that exactly~$s$ nodes are
reachable along the edge connecting~$i$ and~$j$, $\mathcal{N}_i$ is the set
of immediate network neighbors of node~$i$, and $\delta(a,b)$ is the
Kronecker delta.

We now introduce a probability generating function $G_i(z) =
\sum_{s=1}^\infty \pi_i(s)\,z^s$, whose value is given by
\begin{align}
G_i(z) &= \sum_{s=1}^\infty z^s \!\!\sum_{\set{s_j:j\in\mathcal{N}_i}}
          \Biggl[ \prod_{j\in\mathcal{N}_i} \pi_{i\from j}(s_j) \Biggr]
  \,\delta\biggl( s-1, \sum_{j\in\mathcal{N}_i} s_j\biggr)
  \nonumber\\
       &= z \prod_{j\in\mathcal{N}_i}\>\sum_{s_j=0}^\infty
            \pi_{i\from j}(s_j)\,z^{s_j},
\end{align}
which can be conveniently written as
\begin{equation}
G_i(z) = z \prod_{j\in\mathcal{N}_i} H_{i\from j}(z),
\label{eq:giz}
\end{equation}
where $H_{i\from j}(z) = \sum_{s=0}^\infty \pi_{i\from j}(s)\,z^s$ is the
generating function for~$\pi_{i\from j}(s)$.

To calculate $H_{i\from j}(z)$, we note that $\pi_{i\from j}(s)$ is zero if
the edge between $i$ and $j$ is unoccupied (which happens with
probability~$1-p$) and nonzero otherwise (probability~$p$), which means
that $\pi_{i\from j}(0) = 1-p$, and for $s\ge1$
\begin{equation}
\pi_{i\from j}(s) = p \!\!\!\sum_{\set{s_k:k\in\mathcal{N}_j\backslash i}}
                    \Biggl[ \prod_{k\in\mathcal{N}_j\backslash i}\>
                    \!\!\pi_{j\from k}(s_k) \Biggr]
  \delta\biggl( s-1,\!\!\sum_{k\in\mathcal{N}_j\backslash i} s_k\biggr),
\end{equation}
where the notation $\mathcal{N}_j\backslash i$ denotes the set of neighbors
of~$j$ excluding~$i$.  Substituting this expression into the definition of
$H_{i\from j}(z)$ above, we then find that
\begin{equation}
H_{i\from j}(z) = 1 - p + pz \prod_{k\in\mathcal{N}_j\backslash i}
                                   H_{j\from k}(z).
\label{eq:messages}
\end{equation}
This self-consistent equation for the generating function~$H_{i\from j}(z)$
suggests a message-passing algorithm: for any value of~$z$ one guesses (for
instance at random) an initial set of values for the~$H_{i\from j}$ and
feeds them into the right-hand side of Eq.~\eqref{eq:messages}, giving a
new set of values on the left.  Repeating this process to convergence gives
a solution for the generating functions, which can then be substituted into
Eq.~\eqref{eq:giz} to give the generating function for the cluster
probabilities~$\pi_i(s)$, from which we can recover the probabilities
themselves by differentiating.

As an example application of this method, note that, since $\pi_i(s)$ is
the probability that $i$ belongs to a small (non-percolating) cluster of
size~$s$, the probability that it belongs to a small cluster of any size is
$\sum_s \pi_i(s) = G_i(1)$ and the probability that it belongs to the
percolating cluster is one minus this.  Then the expected fraction~$S$ of
the network occupied by the entire percolating cluster is given by the
average over all nodes:
\begin{equation}
S = {1\over n} \sum_{i=1}^n \bigl[ 1 - G_i(1) \bigr]
  = 1 - {1\over n} \sum_{i=1}^n \prod_{j\in\mathcal{N}_i} H_{i\from j}(1).
\label{eq:g1}
\end{equation}
Setting $z=1$ in Eq.~\eqref{eq:messages} we have
\begin{equation}
H_{i\from j}(1) = 1 - p + p \prod_{k\in\mathcal{N}_j\backslash i}
                                   H_{j\from k}(1),
\label{eq:h1}
\end{equation}
and the solution of this equation, for instance by iteration from a random
initial guess, allows us to calculate the size of the percolating
cluster~\cite{Rivoire05}.  We give illustrative applications to several
networks below.

As another example, consider the case were vertex~$i$ does not belong to
the percolating cluster.  Then the expected size~$\av{n_i}$ of the cluster
it does belong to is given by
\begin{equation}
\av{n_i} = {\sum_s s \pi_i(s)\over\sum_s \pi_i(s)}
  = {G_i'(1)\over G_i(1)}
  = 1 + \sum_{j\in\mathcal{N}_j} {H'_{i\from j}(1)\over H_{i\from j}(1)},
\label{eq:avs}
\end{equation}
and, differentiating Eq.~\eqref{eq:messages}, we have
\begin{equation}
H'_{i\from j}(1) = p \biggl[ 1 + \sum_{k\in\mathcal{N}_j\backslash i}
                     {H'_{j\from k}(1)\over H_{j\from k}(1)} \biggr]
               \prod_{k\in\mathcal{N}_j\backslash i} H_{j\from k}(1).
\label{eq:derivative}
\end{equation}
By iterating both Eq.~\eqref{eq:h1} and Eq.~\eqref{eq:derivative} from
random initial values and substituting the results into Eq.~\eqref{eq:avs}
we can calculate the expected cluster size.  Or we can average over all
vertices to calculate the network-wide average size of a non-percolating
cluster.

\begin{figure}
\begin{center}
\includegraphics[width=\columnwidth]{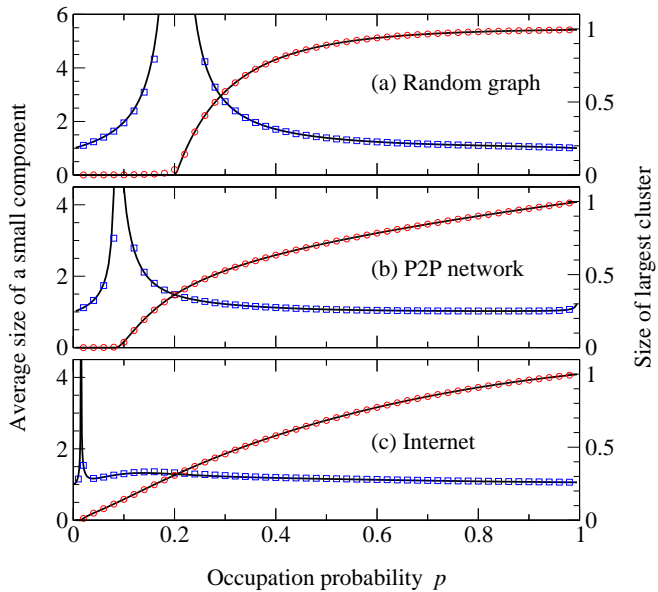}
\end{center}
\caption{(Color online) Results for three example networks.  Solid lines
  are the message passing calculations and the points, included for
  comparison, are from direct numerical simulations on the same networks.
  For each network we show the average size of small clusters (blue
  squares) and the size of the percolating cluster (red circles).  The
  simulations are averaged over at least 100 repetitions of the percolation
  process in each case.  The networks are (a)~an Erd\H{o}s--R\'enyi random
  graph of $10\,000$ nodes and mean degree~5, (b)~a peer-to-peer file
  sharing network of $62\,586$ nodes~\cite{IRF02}, and (c)~a $22\,963$-node
  snapshot of the structure of the Internet at the level of autonomous
  systems.}
\label{fig:results}
\end{figure}

Figure~\ref{fig:results} shows results from the application of these
techniques to the calculation of cluster sizes for three networks: a
computer-generated network which is genuinely tree-like (so the method
should work well), and two real-world networks for which percolation could
be useful as a model of resilience---a network representation of the
Internet at the level of autonomous systems and a peer-to-peer file sharing
network.  Also shown on the figure are results from direct numerical
simulations of percolation on the same networks.  As the figure shows, the
message passing and numerical results are in excellent agreement, not only
for the computer-generated example but also for the two real-world
networks, even though the latter are not tree-like.  Both the size of the
percolating cluster and the mean size of the small clusters are given
accurately by the message passing method.

One might ask what the virtue of the message passing method is if one can
perform direct percolation simulations of the kind used in
Fig.~\ref{fig:results}.  There are two answers to this question.  First,
simulation algorithms calculate cluster sizes for only a single realization
of the randomness inherent in the percolation process.  To get accurate
results the simulation must be repeated over many realizations, but this
can take a significant amount of time and even then the final results still
contain statistical errors.  The message passing method on the other hand
returns the cluster size distribution over all realizations of the
randomness in a single calculation.

Second, and perhaps more intriguing, the message passing method not only
provides a numerical algorithm for percolation calculations but also allows
us to derive new fundamental results by analyzing the behavior of the
algorithm itself.  As an example, we can calculate the exact position of
the percolation threshold on an arbitrarily large, locally tree-like
network, as follows.

The value $H_{i\from j}(1)=1$ for all $i,j$ is trivially a solution of
Eq.~\eqref{eq:h1} and hence also a fixed point under the iteration of that
equation.  Since $H_{i\from j}(1)$ is the probability that vertex~$i$ does
not belong to the percolating cluster, this solution corresponds the
situation in which no vertex is in the percolating cluster.  If the
solution is a stable fixed point of Eq.~\eqref{eq:h1}, then iteration will
converge to it and our message passing algorithm will tell us there is no
percolating cluster.  If it is unstable, we will end up at a different
solution and there is a percolating cluster.  Thus the point at which the
trivial fixed point $H_{i\from j}(1)=1$ goes from being stable to being
unstable is precisely the percolation threshold.

We can determine the stability of the fixed point by linearizing: we write
$H_{i\from j}(1)=1-\epsilon_{i\from j}$ and expand Eq.~\eqref{eq:h1} to
leading order in~$\epsilon_{i\from j}$, which gives $\epsilon_{i\from j} =
p\sum_{k\in\mathcal{N}_j\backslash i} \epsilon_{j\from k}$, or in matrix
notation
\begin{equation}
\boldsymbol{\epsilon} = p \mat{B}\boldsymbol{\epsilon},
\label{eq:linearlized}
\end{equation}
where $\boldsymbol{\epsilon}$ is the $2m$-element vector with
elements~$\epsilon_{i\from j}$ and $\mat{B}$ is a $2m\times2m$
non-symmetric matrix with rows and columns indexed by directed edges
$i\from j$ and elements $B_{i\from j,k\from l} =
\delta_{jk}(1-\delta_{il})$.  This matrix is known as the Hashimoto or
non-backtracking matrix and has been a focus of recent attention for its
role in community detection and centrality calculations on
networks~\cite{Krzakala13,MZN14}.

The vector $\boldsymbol{\epsilon}$ tends to zero and hence the fixed point
is stable under iteration of~\eqref{eq:linearlized} if and only if $p$
times the leading eigenvalue of~$\mat{B}$ is less than unity.  Hence we
conclude that \textit{the critical percolation probability~$p_c$ of a
  sparse, locally tree-like network is equal to the reciprocal of the
  leading eigenvalue of the non-backtracking matrix.}

A different result, reminiscent of this one, has been given recently by
Bollob\'as~\etal~\cite{BBCR10}, who show that in the limit of large network
size the critical occupation probability for percolation on a \defn{dense}
network is equal to the reciprocal of the leading eigenvalue of the
adjacency matrix.  The result given here is the equivalent for sparse
networks.

As a simple example consider a random $k$-regular graph, i.e.,~a network in
which every node has exactly $k$ edges but connections are otherwise made
at random.  For such a graph the non-backtracking matrix has $k-1$ nonzero
elements in each row and column and hence its largest eigenvalue is exactly
$k-1$, giving $p_c = 1/(k-1)$, which can easily be confirmed to be the
correct answer using other methods~\cite{CEBH00,CNSW00}.  The leading
eigenvalue of the adjacency matrix on the other hand, which gives the
dense-graph percolation threshold as discussed above, is $k$ and hence
would give a lower, and incorrect, result of~$p_c=1/k$.

In fact, the leading eigenvalue of the adjacency matrix is never less than
the leading eigenvalue of the non-backtracking matrix.  To see this,
consider a matrix~$\mat{B}'$, which is a slight variant of the
non-backtracking matrix having elements $B'_{i\from j,k\from l} =
\delta_{jk}$.  An eigenvector~$\vec{v}$ of this matrix with elements
$v_{i\from j}$ and eigenvalue~$\lambda$ satisfies
\begin{equation}
\lambda v_{i\from j} = \sum_{k\from l} \delta_{jk} v_{k\from l}
  = \sum_{kl} A_{kl} \delta_{jk} v_{k\from l} = \sum_l A_{jl} v_{j\from l},
\end{equation}
which has solutions $v_{i\from j} = w_j$ where $w_j$ are the elements of
any eigenvector~$\vec{w}$ of the adjacency matrix~$\mat{A}$ and $\lambda$
is the corresponding eigenvalue.  Thus~$\mat{B}'$ and $\mat{A}$ have the
same eigenvalues and in particular the leading eigenvalue of $\mat{B}'$ is
also the leading eigenvalue~$\lambda_A$ of~$\mat{A}$.

We now observe that the difference $\Delta\mat{B} = \mat{B}'-\mat{B}$ has
elements $\Delta B_{i\from j,k\from l} = \delta_{jk}\delta_{il}$, which are
all nonnegative, and we apply the so-called Collatz--Wielandt formula, a
corollary of the Perron--Frobenius theorem which says that for any real
vector~$\vec{x}$ the leading eigenvalue of~$\mat{B}'$ (which, as we have
said, is equal to~$\lambda_A$) satisfies
\begin{equation}
\lambda_A \ge \min_{i:x_i\ne0} \frac{[\mat{B}'\vec{x}]_i}{x_i}
  = \min_{i:x_i\ne0} \biggl( \frac{[\mat{B}\vec{x}]_i}{x_i}
    + \frac{[\Delta\mat{B}\vec{x}]_i}{x_i} \biggr).
\label{eq:cw}
\end{equation}
Let us choose $\vec{x}$ to be the leading eigenvector of~$\mat{B}$, which
has all elements nonnegative by the Perron--Frobenius theorem.  Then
$[\mat{B}\vec{x}]_i/x_i = \lambda_B$, where $\lambda_B$ is the leading
eigenvalue of~$\mat{B}$, and $[\Delta\mat{B}\vec{x}]_i/x_i\ge0$ for
all~$i$, so~\eqref{eq:cw} implies that $\lambda_A\ge\lambda_B$.

This in turn implies that \textit{the dense-matrix result for the
  percolation threshold based on the adjacency matrix is a lower bound on
  the percolation threshold of a sparse tree-like graph.}

An interesting special case is that of a perfect tree, a network with no
loops at all.  Percolation, in the sense of a percolating cluster that
fills a nonzero fraction of the network in the large-$n$ limit, never
occurs on such a network---for all $p<1$ the largest cluster occupies only
a vanishing fraction of the network and our formalism gives this result
correctly.  The diagonal elements of powers of the non-backtracking matrix
count numbers of closed non-backtracking walks on a
graph~\cite{ABLS07,Krzakala13} (hence the name ``non-backtracking
matrix''), but a perfect tree has no such walks, so the trace of every
power of the matrix is zero and hence so also are all eigenvalues.  Thus
the reciprocal of the largest eigenvalue diverges and there is no
percolation threshold.  The leading eigenvalue of the adjacency matrix, on
the other hand, is nonzero on a tree.  On a $k$-regular tree, for instance,
the leading eigenvalue of the adjacency matrix for large~$n$ is~$k$ again,
implying a percolation threshold of~$1/k$.  This is, indeed, a lower bound
on the true percolation threshold, as it must be, but it is in error by a
wide margin.

All of our results so far have been for tree-like networks, but most
real-world networks are not trees.  We can nonetheless use the techniques
developed here to say something about the non-tree-like case.  On a tree
the number of nodes reachable along the edge from~$i$ to~$j$ is one (for
node~$j$ itself) plus the sum of the numbers~$n_{j\from k}$ reachable along
every other edge attached to~$j$.  On a non-tree, on the other hand, this
sum overestimates the number of reachable nodes because some nodes are
reachable along more than one edge from~$j$.  This means that for $z\le1$
the generating function~$H_{i\from j}(z)$ for the true number of reachable
nodes will be greater than or equal to the value given by a naive estimate
calculated from a simple average over the randomness:
\begin{align}
&\quad H_{i\from j}(z) \ge 1 - p + p z \Bigl\langle 
               z^{\sum_{k\in\mathcal{N}_j\backslash i} n_{j\from k}}
  \Bigr\rangle \nonumber\\
  &= 1 - p + pz \biggl\langle \prod_{k\in\mathcal{N}_j\backslash i}\!
                z^{n_{j\from k}} \biggr\rangle
  \ge 1 - p + pz \!\prod_{k\in\mathcal{N}_j\backslash i}\!
                  \bigl\langle z^{n_{j\from k}} \bigr\rangle,
\end{align}
where the second inequality follows by an application of the Chebyshev
integral inequality~\cite{KN10a}.  But $\bigl\langle z^{n_{j\from k}}
\bigr\rangle = H_{j\from k}(z)$ by definition, so we find that on a
non-tree-like network the exact equality of Eq.~\eqref{eq:messages} is
replaced with an inequality:
\begin{equation}
H_{i\from j}(z) \ge 1 - p + pz \prod_{k\in\mathcal{N}_j\backslash i}
                    H_{j\from k}(z).
\label{eq:hineq}
\end{equation}
Suppose, however, that we nonetheless decide to use the exact equality
of~\eqref{eq:messages}, iterating to estimate the generating functions.  If
we start from an initial value of $H_{i\from j}$ equal to the true answer
we are looking for (which we don't know, but let us suppose momentarily
that we do), then it is straightforward to see from~\eqref{eq:hineq} that
the value of $H_{i\from j}$ will never increase under the iteration,
implying that the value we calculate will be a lower bound on the true
value for all~$z\le1$.  As we approach the percolation threshold from above
in the large size limit, the true value of $H_{i\from j}(1)$, which
represents the probability that the edge from~$i$ to~$j$ connects to a
small cluster, approaches~1, while the value calculated from
Eq.~\eqref{eq:messages}, which is less than or equal to the true value,
must reach 1 later, i.e.,~at a lower or equal value of~$p$.  Thus the
percolation threshold estimated from~\eqref{eq:messages} is never higher
than the true percolation threshold.  Equivalently, we can say that
\textit{for any network, $p_c$~is always greater than or equal to the
  inverse of the leading eigenvalue of the non-backtracking matrix}.  The
only exception is for the case of a perfect tree, for which the largest
eigenvalue is zero, as discussed above.  Thus the leading eigenvalue gives
us a bound on the percolation threshold.

We can also combine this result with our earlier observation that the
leading eigenvalue of the adjacency matrix is never less than that of the
non-backtracking matrix to make the further statement that
\textit{$p_c$~for any network is always greater than or equal to the
  inverse of the leading eigenvalue of the adjacency matrix.}  Thus, both
eigenvalues place lower bounds on~$p_c$, but the bound given by the
non-backtracking matrix is better (or at least never worse) than the one
given by the adjacency matrix.  Numerical tests of these results on various
networks are given in the Supplemental Information.

In summary, we have in this paper shown that percolation on sparse, locally
tree-like networks can be reformulated as a message passing process,
allowing us to solve for average percolation properties such as the size of
the percolating cluster and the average size of the non-percolating
clusters.  Tests on both computer generated and real-world networks show
good agreement with numerical simulations of percolation on the same
networks.  By analyzing the message passing equations we have also shown
that the position of the percolation threshold on tree-like networks is
given by the inverse of the leading eigenvalue of the non-backtracking
matrix.  On non-tree-like networks this result is not exact but it gives a
bound on the exact result.

\pagebreak The authors thank Cris Moore, Leonid Pryadko, and Pan Zhang for
useful conversations.  After this work was completed we learned of
concurrent work by Hamilton and Pryadko~\cite{HP14} in which a similar
result for the percolation threshold is derived.  This work was funded in
part by the National Science Foundation under grants DMS--1107796 and
DMS--1407207 and by DARPA under grant FA9550--12--1--0432.

\appendix
\section{Supplemental information}
\subsection{Numerical calculation of the leading eigenvalue}
One can calculate the leading eigenvalue of the non-backtracking matrix
numerically and invert to determine the percolation threshold, but the
matrix has size $2m\times2m$, which can become quite large, making the
calculation cumbersome.  It can be sped up by using the so-called Ihara (or
Ihara-Bass) determinant formula as described in~\cite{Krzakala13}, where it
is shown that the leading eigenvalue of the non-backtracking matrix is also
the leading eigenvalue of the $2n\times2n$ matrix
\begin{equation}
\setlength{\arraycolsep}{6pt}
\mat{M} = \begin{pmatrix}
             \mat{A} & \mat{I}-\mat{D} \\
             \mat{I} & \mat{0}
          \end{pmatrix},
\end{equation}
where $\mat{D}$ is the diagonal matrix with the node degrees along its
diagonal.  For a sparse network this matrix is also sparse, with only
$2m+2n$ nonzero elements---far fewer than the non-backtracking matrix
itself---which permits rapid numerical calculation of the leading
eigenvalue.  This method was used to calculate the values given in the
following section.

\subsection{Percolation thresholds}
We have shown that the inverse leading eigenvalues of the adjacency matrix
and the non-backtracking matrix both provide lower bounds on the
percolation threshold on a sparse network, but that the non-backtracking
matrix always gives a better bound (or at least no worse).  Moreover, on a
network that is locally tree-like the non-backtracking matrix gives the
exact threshold.

Table~\ref{tab:results} shows percolation thresholds, both estimated and
measured, for a range of sparse networks.  For each network we have
computed an approximation to the true percolation threshold by repeated
numerical simulations and the bounds given by the leading eigenvalues of
the non-backtracking and adjacency matrices.

On regular lattices, the common method for calculating the position of the
percolation threshold is to look for the point at which a cluster forms
that spans the lattice from edge to edge, but this is not possible on a
network since a network has no edges.  Instead, therefore, we identify the
percolation threshold by looking at the size of the second-largest cluster.
The largest cluster has size that always increases with increasing~$p$, but
the second-largest peaks at the percolation threshold and then falls off
again, so the point of largest size can be used as an estimate of the
position of the threshold.

As the table shows, the results for the percolation threshold are in good
agreement for the two computer-generated networks (the random graph and the
block model), which are genuinely tree-like.  The four remaining networks
on the other hand are not tree-like and hence we don't expect exact
agreement and this is confirmed by the results in the table.  The degree of
disagreement varies from case to case, but in all cases the
non-backtracking matrix gives a lower bound on the true threshold, and it
gives a better bound than the adjacency matrix.

\begin{table}
\setlength{\tabcolsep}{4pt}
\begin{tabular}{l|lll}
        & \multicolumn{3}{c}{Percolation threshold} \\
Network & Adjacency & Non-backtracking & Actual \\
\hline
Random graph & 0.161  & 0.200  & 0.200     \\
Block model  & 0.140  & 0.167  & 0.173(1)  \\
Circuit	     & 0.200  & 0.340  & 0.465(2)  \\
Gnutella     & 0.0759 & 0.0871 & 0.0967(2) \\
Internet     & 0.0140 & 0.0155 & 0.0231(1) \\
Amazon       & 0.0426 & 0.0562 & 0.097(1)
\end{tabular}
\caption{Percolation thresholds estimated from the eigenvalues of the
  adjacency and non-backtracking matrices, and measured directly in numerical
  simulations (or calculated exactly in the case of the random graph).  The
  networks are: a Poisson random graph with average degree~5 and $100\,000$
  nodes; a stochastic block model with $100\,000$ nodes, four groups, and
  an average of 4 in-group and 2 out-group edges per node; electronic
  circuit 838 from the ISCAS 89 benchmark set~\cite{Milo04b}; a snapshot of
  the Internet autonomous system peering structure; a Gnutella peer-to-peer
  filesharing network~\cite{IRF02}; and a copurchasing network of items on
  Amazon.com~\cite{LAH07}.  The numerical estimates of~$p_c$ are obtained
  by finding the point at which the size of the second-largest cluster is
  greatest.  Figures in parentheses indicate the error on the last digit.}
\label{tab:results}
\end{table}

\end{document}